\documentclass[aps,prl,preprint]{revtex4}

\usepackage{epsfig}

\draft

\begin{document}
\author{Hong-shi Zong$^{1,2}$ and Wei-min Sun$^{1,2}$}
\address{$^{1}$ Department of Physics, Nanjing University, Nanjing 210093, China}
\address{$^{3}$ Joint Center for Particle, Nuclear Physics and Cosmology, Nanjing 210093, China}

\title{The calculation of the equation of state of QCD at finite chemical potential and zero temperature}

\begin{abstract}
In this paper, we give a direct method for calculating the partition function, and hence the equation of state (EOS) of Quantum Chromodynamics (QCD) at finite chemical potential and zero temperature. In the EOS derived in this paper the pressure density is the sum of two terms: the first term ${\cal P}(\mu)|_{\mu=0}$ (the pressure density at $\mu=0$) is a $\mu$-independent constant; the second term, which is totally determined by $G_R[\mu](p)$ (the renormalized dressed quark propagator at finite $\mu$), contains all the nontrivial $\mu$-dependence. By applying a general result in the rainbow-ladder approximation of the Dyson-Schwinger approach obtained in our previous study [Phys. Rev. {\bf C 71}, 015205 (2005)], $G_R[\mu](p)$ is calculated from the meromorphic quark propagator proposed in [Phys. Rev. {\bf D 70}, 014014 (2004)]. From this the full analytic expression of the EOS of QCD at finite $\mu$ and zero $T$ is obtained (apart from the constant term ${\cal P}(\mu)|_{\mu=0}$ which can in principle be calculated from the CJT effective action). A comparison between our EOS and the cold, perturbative EOS of QCD of Fraga, Pisarski and Schaffner-Bielich is made. It is expected that our EOS can provide a possible new approach for the study of neutron stars.

\bigskip


\bigskip

E-mail: zonghs@chenwang.nju.edu.cn.

\bigskip


\end{abstract}

\maketitle
The study of the partition function is at the crux of equilibrium statistical field theory (see, for example, Refs. [1,2]). The thermodynamic properties of a system, hence also the equation of state (EOS), are completely determined by the partition function. The calculation of the partition function of Quantum Chromodynamics (QCD) at finite chemical potential is a contemporary focus; e.g., Refs. [3-9]. In addition, it is well-known that in astrophysics the study of the neutron star depends crucially on the assumed EOS [10,11]. The study of EOS of QCD is thus of extreme importance. In this paper, we try to give a direct method for calculating the partition function and EOS of QCD at finite chemical potential.

The renormalized partition function of QCD at zero temperature and finite chemical potential reads
\begin{eqnarray}
{\cal{Z}}[\mu]&=&\int{\cal{D}}\bar{q_R}{\cal{D}}q_R{\cal{D}}A_R~\exp\left\{-S_{R}[\bar{q}_R,q_R,A_R]+\int d^4x~\mu Z_2\bar{q}_R(x)\gamma_{4}q_R(x)\right\},
\end{eqnarray}
where $S_{R}[\bar{q}_R,q_R,A_R]$ is the standard renormalized
Euclidean QCD action with $q_R$ being the renormalized quark field
with three flavors and three colors, $Z_2=Z_2(\zeta^2,\Lambda^2)$ is
the quark wave-function renormalization constant ($\zeta$ is the
renormalization point and $\Lambda$ is the regularization
mass-scale). Here we leave the ghost field term and its integration
measure to be understood. The pressure density ${\cal{P}}(\mu)$ is
given by
\begin{equation}
{\cal{P}}(\mu)=\frac{1}{{\cal{V}}}~\ln\cal{Z}[\mu],
\end{equation}
where ${\cal{V}}$ is the four-volume normalising factor. The above equation for the pressure is just the EOS and from this one immediately obtains the quark-number density
\begin{eqnarray}
\rho(\mu)&=&\frac{\partial {\cal P}(\mu)}{\partial\mu}=\frac{1}{\cal{V}}\frac{1}{\cal{Z}[\mu]}\frac{\partial \cal{Z}[\mu]}{\partial\mu}\nonumber\\
&=&\frac{1}{{\cal{V}}}\frac{\int{\cal{D}}\bar{q}_R{\cal{D}}q_R{\cal{D}}A_R\int d^4x Z_2\bar{q}_R(x)\gamma_{4}q_R(x)\exp\left\{-S_R[\bar{q}_R,q_R,A_R;\mu]\right\}}{\int{\cal{D}}\bar{q}_R{\cal{D}}
q_R{\cal{D}}A_R~\exp\left\{-S_R[\bar{q}_R,q_R,A_R;\mu]\right\}},
\end{eqnarray}
where $S_R[\bar{q}_R,q_R,A_R;{\mu}]\equiv S_R[\bar{q}_R,q_R,A_R]-\int d^4x~\mu Z_2\bar{q}_R(x)\gamma_{4}q_R(x)$.

On the other hand, the dressed quark propagator at finite chemical potential can be written as
\begin{equation}
G_{Rij}[\mu](x,y)=\frac{\int{\cal{D}}\bar{q}_R{\cal{D}}q_R{\cal{D}}A_R~q_{Ri}(x)\bar{q}_{Rj}(y)\exp\left\{-S_R[\bar{q}_R,q_R,A_R;\mu]\right\}}{\int{\cal{D}}\bar{q}_R{\cal{D}}q_R{\cal{D}}A_R~\exp\left\{-S_R[\bar{q}_R,q_R,A_R;\mu])\right\}}.
\end{equation}
From Eq. (4), it is easy to obtain the following
\begin{equation}
\mathrm{Tr}\left\{G_R[\mu]\gamma_4\right\}=-\frac{\int{\cal{D}}\bar{q}_R{\cal{D}}q_R{\cal{D}}
A_R\int d^4x\bar{q}_R(x)\gamma_{4}q_R(x)\exp\left\{-S_R[\bar{q}_R,q_R,A_R;\mu]\right\}}{\int{\cal{D}}\bar{q}_R{\cal{D}}q_R{\cal{D}}A_R~\exp\left\{-S_R[\bar{q}_R,q_R,A_R;\mu]\right\}},
\end{equation}
where the notation $\mathrm{Tr}$ denotes trace over the color, flavor, Dirac and coordinate space indices. Comparing Eq. (3) with (5), we obtain a well-known result (for its recent application, see, e.g. Ref. [12])
\begin{eqnarray}
\rho(\mu)&=&-\frac{Z_2}{{\cal{V}}}Tr\left\{G_R[\mu]\gamma_4\right\}=-N_cN_f Z_2 \int\frac{d^4p}{(2\pi)^4}\mathrm{tr}\left\{G_R[\mu](p)\gamma_4\right\},
\end{eqnarray}
where $N_c$ and $N_f$ denote the number of colors and of flavors, respectively, and the trace operation is over Dirac indices. From Eq. (6) it can be seen that the quark-number density $\rho(\mu)$ is totally determined by the dressed quark propagator at finite chemical potential. Setting $\mu=0$ in Eq. (6), one finds that the quark-number density at zero chemical potential vanishes. This is because by writing out the general Lorentz structure of $G_R[\mu=0](p)$ and performing the trace one can verify that the integrand is an odd function of $p_4$ and therefore the integration vanishes. This is what one expects in advance.

Integrating the equation $\rho(\mu)=\frac{\partial {\cal P}(\mu)}{\partial\mu}$, one obtains
\begin{equation}
{\cal{P}}(\mu)=\left.{\cal{P}}(\mu)\right|_{\mu=0}+\int_{0}^{\mu}d\mu'\rho(\mu')=\left.{\cal{P}}(\mu)\right|_{\mu=0}-N_cN_fZ_2\int_{0}^{\mu}d\mu'\int\frac{d^4p}{(2\pi)^4}
\mathrm{tr}\left\{G_R[\mu'](p)\gamma_4\right\}.
\end{equation}
From the above equation it can be seen that the pressure density ${\cal P}(\mu)$ is the sum of two terms: the first term ${\cal P}(\mu)|_{\mu=0}$ (the pressure density at zero $\mu$) is only a $\mu$-independent constant; the second term, which is totally determined by $G_R[\mu](p)$, contains all the nontrivial $\mu$-dependence. Here we note that formula (7) is formally model-independent. However, at present it is very difficult to calculate ${\cal P}(\mu)|_{\mu=0}$ and $G_R[\mu](p)$ from first principles of QCD. So when one uses formula (7) to calculate the EOS of QCD, one has to resort to various nonperturbative QCD models.

Over the past few years, considerable progress has been made in the framework of the rainbow-ladder approximation of the Dyson-Schwinger (DS) approach [13-16], which provides a successful description of various nonperturbative aspects of strong interaction physics. We naturally expect that it might be a useful nonperturbative approach in the study of EOS of QCD at finite chemical potential. In this paper we shall employ this approach.

Now let us turn to the actual calculation of $G_R[\mu](p)$. Here we apply the following general result proved in Refs. [17,18]: Under the rainbow approximation of the Dyson-Schwinger equation (DSE), if one ignores the
$\mu$ dependence of the dressed gluon propagator (this is a commonly used approximation in calculating the dressed quark propagator at finite chemical potential [14,17-23]) and assumes that the dressed quark propagator at finite $\mu$ is analytic in the neighborhood of $\mu=0$, then the inverse dressed quark propagator at finite chemical potential can be obtained from the one at zero chemical potential by the following simple substitution [17,18]:
\begin{equation}
G_R^{-1}[\mu](p)=G_R^{-1}({\tilde p})=i \gamma \cdot {\tilde p} A({\tilde p}^2)+B({\tilde p}^2),
\end{equation}
where $\tilde{p}=(\vec{p},p_4+i\mu)$ and $G_R^{-1}(p)=i \gamma \cdot p A( p^2)+B(p^2)$ is the inverse dressed quark propagator at $\mu=0$. 
Here one may ask whether the quark propagator given by the above equation is valid for any $\mu$. This question has two levels, those of mathematics and physics. On the mathematical level, it should be noted that in deriving Eq. (8) we have assumed that $|\mu|$ is smaller than the radius of convergence of Taylor expansion of $G_R^{-1}[\mu](p)$ around $\mu=0$ (for details, see Ref. [17]). However, Eq. (8) holds in the whole domain of analyticity of $G_R^{-1}[\mu](p)$ in the complex $\mu$-plane, not only within the circle of convergence of $\mu$ expansion. This is a result of a well-known theorem in complex analysis [24]: Suppose each of two functions $f(z)$ and $g(z)$ is analytic in a common domain $D$. If $f(z)$ and $g(z)$ coincide in some subportion $D'\subset D$, then $f(z)=g(z)$ everywhere in $D$. On the physical level, it is generally believed that at finite $\mu$ there will be a chiral transition. In particular, in the chiral limit, this is a first order phase transition, and hence the quark mass function derived from the quark propagator will drop discontinuously to zero. This feature is not reproduced by the quark propagator in Eq. (8). Hence it should be stressed that this not happening is a pure assumption of the model employed in this paper.

According to Eq. (8), once the dressed quark propagator
at $\mu=0$ is known, one can obtain the dressed quark propagator at finite $\mu$ by means of Eq. (8). Therefore, in order to calculate $G_R[\mu](p)$ using Eq. (8), one needs to specify the form of the dressed quark propagator at zero chemical potential in advance. In Ref. [25], guided by the solution of the coupled set of DSEs for the ghost, gluon and quark propagator in the Landau gauge, the following meromorphic form of the dressed quark propagator is proposed:
\begin{equation}
G_R(p)=Z_2^{-1}(\zeta^2,\Lambda^2)\sum_{j=1}^{n_P} \left\{\frac{r_j}{i \not\! p +m_j}+\frac{r_j}{i \not\! p+m_j^*}\right\},
\end{equation}
with $m_j=a_j+ib_j$. The propagator of this form has $n_P$ pairs of complex conjugate poles located at $a_j\pm ib_j$. When some $b_j$ is set to zero, the pair of complex conjugate poles degenerates to a real pole. The residues $r_j$ are real (note that a similar meromorphic form of the quark propagator was previously proposed in Ref. [26], in which the residues in the two additive terms are complex conjugate of each other).  In the chiral limit, the requirement that the dressed quark propagator reduces to the free one in the large momentum limit entails that 
\[
\sum_{j=1}^{n_P}\,r_j=\frac{1}{2}~~~ \mbox{and} ~~~\sum_{j=1}^{n_P}\,r_ja_j=0.
\]
In this paper, following Ref. [25], we set the renormalization point to be $\zeta^2=16~\mathrm{GeV}^2$. 
Here it should be noted that the quark propagator (9) is obtained from a calculation going significantly beyond the rainbow approximation (for details, see Ref. [25]) and not necessarily a solution to the rainbow DSE for the quark propagator at $\mu=0$. However, it would be valid to assume for the moment that the quark propagator obtained by substituting (9) into Eq. (8) is an acceptable approximation for the quark propagator at finite chemical potential and see what physical results it will yield.
With this in mind, from the form of $G_R(p)$ given in Eq. (9) we obtain
\begin{eqnarray}
G_R[\mu](p)&=&Z_2^{-1}(\zeta^2,\Lambda^2)\sum_{j=1}^{n_P}\left\{\frac{r_j}{i \not\! {\tilde p}+m_j}+\frac{r_j}{i \not\! {\tilde p}+m_j^*}\right\} \nonumber \\
&=& Z_2^{-1}(\zeta^2,\Lambda^2)\sum_{j=1}^{n_P} \left\{\frac{r_j (-i \not\! {\tilde p}+m_j)}{{\tilde p}^2+m_j^2}+\frac{r_j (-i \not\! {\tilde p}+m_j^*)}{{\tilde p}^2+m_j^{*2}}\right\}.
\end{eqnarray}
Substituting the above equation into Eq. (6) and performing the trace, one obtains
\begin{eqnarray}
\rho(\mu)&=& 4i N_c N_f \int \frac{d^4 p}{(2\pi)^4}\sum_{j=1}^{n_P}\left\{\frac{r_j(p_4+i\mu)}{{\vec p}^2+(p_4+i\mu)^2+m_j^2}+\frac{r_j(p_4+i\mu)}{{\vec p}^2+(p_4+i\mu)^2+m_j^{*2}}\right\} \\
&=& \frac{4i N_c N_f}{(2\pi)^4}\sum_{j=1}^{n_P}\left\{r_j \int d^4 p \frac{p_4+i\mu}{{\vec p}^2+(p_4+i\mu)^2+m_j^2}+r_j \int d^4 p \frac{p_4+i\mu}{{\vec p}^2+(p_4+i\mu)^2+m_j^{*2}}\right\}.\nonumber
\end{eqnarray}
Now we need to evaluate the integrals in Eq. (11). We first consider the first integral on the right-hand-side of (11) and write
\[
\int d^4 p \frac{p_4+i\mu}{{\vec p}^2+(p_4+i\mu)^2+m_j^2}=\int d {\vec p} \int\limits_{-\infty}^{+\infty}d p_4 \frac{p_4+i\mu}{{\vec p}^2+(p_4+i\mu)^2+m_j^2}.
\]
The integral over $p_4$ can be written as a line integral from $-\infty+i\mu$ to $+\infty+i\mu$ in the complex plane:
\[
\int\limits_{-\infty}^{+\infty}d p_4 \frac{p_4+i\mu}{{\vec p}^2+(p_4+i\mu)^2+m_j^2}=\int\limits_{-\infty+i\mu}^{+\infty+i\mu}dz \frac{z}{z^2+{\vec p}^2+m_j^2}.
\]
The latter integral can be evaluated by means of contour integration. For this purpose we choose the following contour (see Fig. 1). The function $f(z)=\frac{z}{z^2+{\vec p}^2+m_j^2}$ has two poles in the complex plane which are located at
\[
z_j=\pm \sqrt{\frac{\sqrt{({\vec p}^2+\alpha_j)^2+\beta_j^2}-{\vec p}^2-\alpha_j}{2}}\pm i~ sgn(-\beta_j)\sqrt{\frac{\sqrt{({\vec p}^2+\alpha_j)^2+\beta_j^2}+{\vec p}^2+\alpha_j}{2}},
\]
where $m_j^2 \equiv \alpha_j+\beta_j i$ and $sgn(-\beta_j)$ denotes the sign of $-\beta_j$.
Among the two poles, one lies in the upper half plane, the other lies in the lower half plane. The imaginary part of the pole in the upper half plane
is
\[
\omega_j ({\vec p})=\sqrt{\frac{\sqrt{({\vec p}^2+\alpha_j)^2+\beta_j^2}+{\vec p}^2+\alpha_j}{2}}.
\]
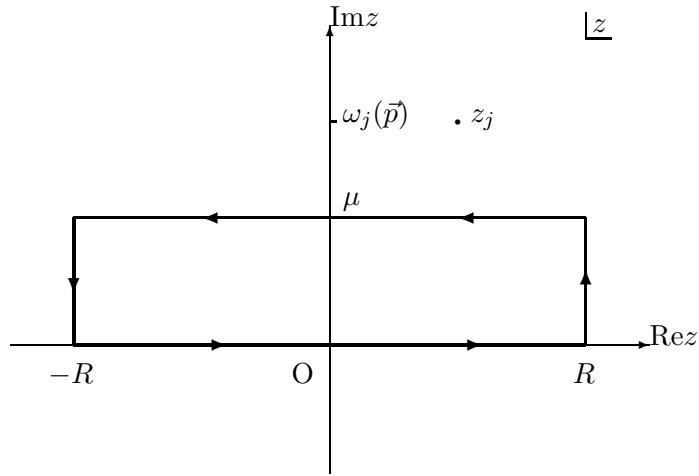
\begin{figure}[b]
\setlength{\unitlength}{1.7cm}
\begin{picture}(5,5)
\put(2.2,2.2){O} \put(5.0,2.5){Re$z$} \put(2.5,5.0){Im$z$}
\put(0.0,2.5){\vector(1,0){5.0}} \put(2.5,1.5){\vector(0,1){3.5}}
\thicklines\put(1.5,3.5){\line(-1,0){1}}
\put(2.5,3.5){\vector(-1,0){1}}\put(3.5,3.5){\line(-1,0){1}}
\put(4.5,3.5){\vector(-1,0){1}} \put(0.5,2.5){\vector(1,0){1.2}}
\put(1.7,2.5){\line(1,0){0.8}}\put(2.5,2.5){\vector(1,0){1.2}}
\put(3.7,2.5){\line(1,0){0.8}}
\put(4.5,2.5){\vector(0,1){0.6}}\put(4.5,3.1){\line(0,1){0.4}}
\put(0.5,3.5){\vector(0,-1){0.6}}\put(0.5,2.9){\line(0,-1){0.4}}
\put(0.3,2.2){$-R$}\put(4.4,2.2){$R$} \put(2.6,3.6){$\mu$}
\put(2.5,4.25){\line(1,0){0.05}}\put(2.6,4.25){$\omega_j(\vec{p})$}
\put(3.5,4.25){\circle*{0.05}}\put(3.6,4.25){$z_j$}
 \thinlines
\put(4.5,4.9){\line(1,0){0.2}} \put(4.5,4.9){\line(0,1){0.2}}
\put(4.55,4.95){$z$}
\end{picture}
\vspace{-3.0cm}
 \caption{The integration contour in the complex $z$
plane}
\end{figure}
Here one should distinguish two cases:\\
(i) $\mu<\omega_j({\vec p})$.
In this case, since there are no poles inside the contour (see Fig. 1), the integral along the contour vanishes. The integral along the straight segment from $-R$ to $R$ vanishes because $f(z)$ is an odd function. In the $R\rightarrow \infty$ limit, the integral along the two vertical segments gives zero contribution because $f(z)$ vanishes at infinity. So in the $R\rightarrow \infty$ limit, the contour integral has its contribution only from $\int\limits_{R+i\mu}^{-R+i\mu} dz f(z)$. Therefore one has
\begin{equation}
\int\limits_{-\infty+i\mu}^{+\infty+i\mu} dz f(z)=0,~~~\mu < \omega_j({\vec p}).
\end{equation}
(ii) $\mu > \omega_j({\vec p})$.
In this case, there is one pole inside the contour (see Fig. 2). According to the above explanation and applying the residue theorem, one has

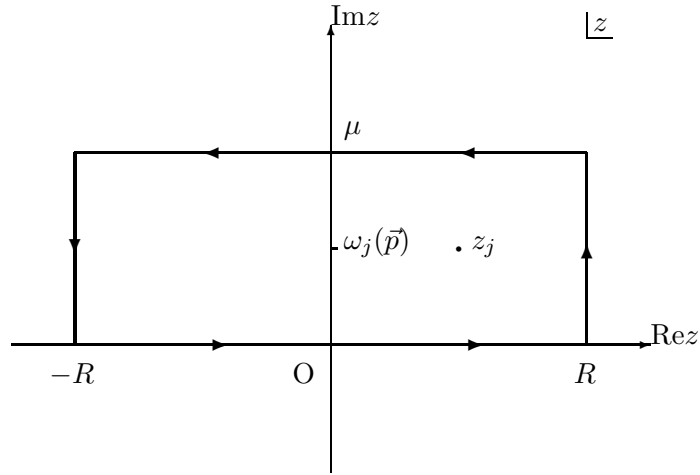
\begin{figure}[b]
\setlength{\unitlength}{1.7cm}
\begin{picture}(5,5)
\put(2.2,2.2){O} \put(5.0,2.5){Re$z$} \put(2.5,5.0){Im$z$}
\put(0.0,2.5){\vector(1,0){5.0}} \put(2.5,1.5){\vector(0,1){3.5}}
\thicklines\put(1.5,4.0){\line(-1,0){1}}
\put(2.5,4.0){\vector(-1,0){1}}\put(3.5,4.0){\line(-1,0){1}}
\put(4.5,4.0){\vector(-1,0){1}} \put(0.5,2.5){\vector(1,0){1.2}}
\put(1.7,2.5){\line(1,0){0.8}}\put(2.5,2.5){\vector(1,0){1.2}}
\put(3.7,2.5){\line(1,0){0.8}}
\put(4.5,2.5){\vector(0,1){0.8}}\put(4.5,3.3){\line(0,1){0.7}}
\put(0.5,4.0){\vector(0,-1){0.8}}\put(0.5,3.2){\line(0,-1){0.7}}
\put(0.3,2.2){$-R$}\put(4.4,2.2){$R$} \put(2.6,4.15){$\mu$}
\put(2.5,3.25){\line(1,0){0.05}}\put(2.6,3.25){$\omega_j(\vec{p})$}
\put(3.5,3.25){\circle*{0.05}}\put(3.6,3.25){$z_j$}
 \thinlines
\put(4.5,4.9){\line(1,0){0.2}} \put(4.5,4.9){\line(0,1){0.2}}
\put(4.55,4.95){$z$}
\end{picture}
\vspace{-3.0cm}
 \caption{The integration contour in the complex $z$ plane}
\end{figure}

\begin{equation}
-\int\limits_{-\infty+i\mu}^{+\infty+i\mu} dz f(z)=2\pi i ~\mathrm{Res}(f(z), z=z_j)=\pi i,~~~\mu > \omega_j({\vec p}).
\end{equation}
Eqs. (12) and (13) can be combined into the following form
\begin{equation}
\int\limits_{-\infty}^{+\infty}d p_4 \frac{p_4+i\mu}{{\vec p}^2+(p_4+i\mu)^2+m_j^2}=\int\limits_{-\infty+i\mu}^{+\infty+i\mu} dz f(z)=-\pi i \theta(\mu-\omega_j({\vec p})).
\end{equation}
Now, when $\mu < \sqrt{\frac{\alpha_j+\sqrt{\alpha_j^2+\beta_j^2}}{2}}$, one has $\omega_j({\vec p}) >\mu, \forall~ {\vec p}$,
so in this case
\[
\theta(\mu-\omega_j({\vec p}))\equiv 0.
\]
When $\mu \geq \sqrt{\frac{\alpha_j+\sqrt{\alpha_j^2+\beta_j^2}}{2}}$, $\omega_j({\vec p}) <\mu$ if and only if $|{\vec p}|< (\mu^2-\frac{\beta_j^2}{4\mu^2}-\alpha_j)^{1/2}$,
so in this case
\[
\theta(\mu-\omega_j({\vec p}))=\left\{
                                     \begin{array}{ll}
                                      1 , & \mathrm{when}~~ |{\vec p}|< (\mu^2-\frac{\beta_j^2}{4\mu^2}-\alpha_j)^{1/2} \\
                                      0, & \mathrm{when}~~ |{\vec p}|>(\mu^2-\frac{\beta_j^2}{4\mu^2}-\alpha_j)^{1/2}.
                                     \end{array}
                               \right.
\]
Therefore one has
\begin{eqnarray}
&&\int d^4 p \frac{p_4+i\mu}{{\vec p}^2+(p_4+i\mu)^2+m_j^2}=-\pi i \int d {\vec p}~ \theta(\mu-\omega_j({\vec p})) \nonumber \\
&&=\left\{\begin{array}{cc}
0,~ & \mathrm{when} ~\mu < \sqrt{\frac{\alpha_j+\sqrt{\alpha_j^2+\beta_j^2}}{2}}  \\
-\frac{4\pi^2 i}{3}(\mu^2-\frac{\beta_j^2}{4\mu^2}-\alpha_j)^{3/2},~ & \mathrm{when} ~\mu \geq \sqrt{\frac{\alpha_j+\sqrt{\alpha_j^2+\beta_j^2}}{2}}
                 \end{array}\right. \nonumber \\
&&=-\frac{4\pi^2 i}{3}\theta(\mu-\sqrt{\frac{\alpha_j+\sqrt{\alpha_j^2+\beta_j^2}}{2}})(\mu^2-\frac{\beta_j^2}{4\mu^2}-\alpha_j)^{3/2}.
\end{eqnarray}
The second integral on the right-hand-side of Eq. (11) can be obtained by making the replacement $m_j^2=\alpha_j+\beta_j i \rightarrow m_j^{*2}=\alpha_j-\beta_j i$ in Eq. (15) and it is equal to the first one:
\begin{equation}
\int d^4 p \frac{p_4+i\mu}{{\vec p}^2+(p_4+i\mu)^2+m_j^{*2}} = -\frac{4\pi^2 i}{3}\theta(\mu-\sqrt{\frac{\alpha_j+\sqrt{\alpha_j^2+\beta_j^2}}{2}})(\mu^2-\frac{\beta_j^2}{4\mu^2}-\alpha_j)^{3/2}.
\end{equation}
Substituting Eqs. (15) and (16) into Eq. (11) gives
\begin{equation}
\rho(\mu)=\frac{2N_c
N_f}{3\pi^2}\sum_{j=1}^{n_P}r_j\theta(\mu-\sqrt{\frac{\alpha_j+\sqrt{\alpha_j^2+\beta_j^2}}{2}})(\mu^2-\frac{\beta_j^2}{4\mu^2}-\alpha_j)^{3/2}.
\end{equation}
In the numerical calculations in this paper, we use three sets of parameters given in Ref. [25], which represent three forms of the propagator: three real poles (3R), two pairs of complex conjugate poles (2CC) and one real pole and one pair of complex conjugate poles (1R1CC). These parameters are listed in Table I.
\vspace{0.2cm}

\begin{center}
\begin{minipage}{16cm}
\begin{center}
{\scriptsize Table I. The parameters used in this paper. They are taken directly from Table II of Ref. [25]}.

\vspace{0.1cm}

\begin{tabular*}{16cm}{l@{\extracolsep{\fill}}*{8}{c}} \hline\hline
Parametrization&$r_1$&$a_1$ (GeV)&$b_1$ (GeV)&$r_2$&$a_2$
(GeV)&$b_2$ (GeV)&$r_3$&$a_3$
(GeV)\\
\hline 2CC&0.360&0.351&0.08&0.140&-0.899&0.463&-&-\\
\hline 1R1CC&0.354&0.377&-&0.146&-0.91&0.45&-&-\\
\hline 3R&0.365&0.341&-&1.2&-1.31&-&-1.06&-1.40\\
\hline\hline
\end{tabular*}
\end{center}
\end{minipage}
\end{center}

The dependence of $\rho(\mu)$ on $\mu$ is plotted in Fig. 3. Just as shown in Fig. 3, the obtained quark-number density distribution differs significantly from the Fermi distribution of the free quark theory. Physically this is a consequence of dynamical chiral symmetry breaking and confinement in the low energy region. We note that when $\mu$ is smaller than a critical value $\mu_0$ ($\mu_0=351~\mathrm{MeV}$, $377~\mathrm{MeV}$ and $341~\mathrm{MeV}$ for the 2CC, 1R1CC and 3R parametrization, respectively), the quark-number density vanishes identically. Namely, $\mu=\mu_0$ is a singularity which separates two regions with different quark number densities. This result agrees qualitatively with the general conclusion of Ref. [27]. In that reference, based on a universal argument, it is pointed out that the existence of some singularity at the point $\mu=\mu_0$ and $T=0$ is a robust and model-independent prediction. The numerical value of the critical chemical potential in pure QCD (i.e., with electromagnetic interactions being switched off) is estimated to be $(m_N-16~\mathrm{MeV})/N_c=307~\mathrm{MeV}$ (where $m_N$ is the nucleon mass and $N_c=3$ is the number of colours). The value of the critical chemical potential obtained in this study is almost of the same order of magnitude as the estimate in that reference. This difference can be attributed to the choice of the parameters of the model quark propagator employed in this paper. In fact, as was pointed out in Ref. [25], the choice of the parameters of the model quark propagator (9) has some arbitrariness. We expect that our work can give further constraints on the model parameters.
\begin{figure}
\epsfig{file=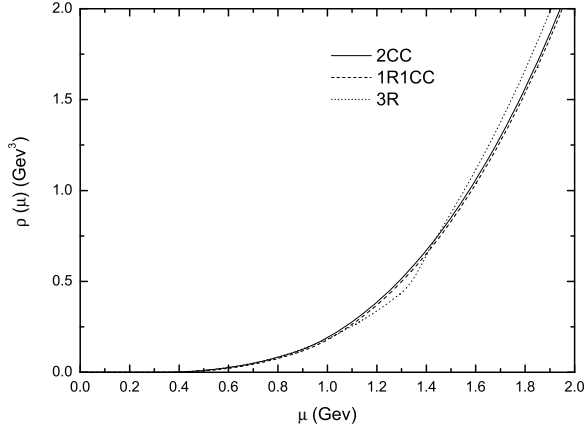,width=9cm}

\caption {Relation between the quark-number density and the chemical potential}
\end{figure}

Now let us calculate $\int\limits_0^\mu d\mu' \rho(\mu')$. From Eq. (17) one obtains
\begin{eqnarray}
\int_0^\mu d\mu' \rho(\mu') &=& \frac{2N_c N_f}{3\pi^2}\sum_{j=1}^{n_P}r_j\int\limits_0^\mu d\mu'\theta \Bigg(\mu'-\sqrt{\frac{\alpha_j+\sqrt{\alpha_j^2+\beta_j^2}}{2}}\Bigg)(\mu'^2-\frac{\beta_j^2}{4\mu'^2}-\alpha_j)^{3/2} \nonumber \\
&=&\frac{2N_c N_f}{3\pi^2}\sum_{j=1}^{n_P}r_j~\theta\Bigg(\mu-\sqrt{\frac{\alpha_j+\sqrt{\alpha_j^2+\beta_j^2}}{2}}\Bigg)\int\limits_{\sqrt{\frac{\alpha_j+\sqrt{\alpha_j^2+\beta_j^2}}{2}}}^\mu d\mu' (\mu'^2-\frac{\beta_j^2}{4\mu'^2}-\alpha_j)^{3/2} \nonumber \\
&=& \frac{2 N_c
N_f}{3\pi^2}\sum_{j=1}^{n_P}r_j~\theta\Bigg(\mu-\sqrt{\frac{\alpha_j+\sqrt{\alpha_j^2+\beta_j^2}}{2}}\Bigg)
I(\mu;\alpha_j,\beta_j).
\end{eqnarray}
where $I(\mu;\alpha_j,\beta_j)$ is:
\begin{eqnarray}
I(\mu;\alpha_j,\beta_j)&\equiv&\int\limits_{\sqrt{\frac{\alpha_j+\sqrt{\alpha_j^2+\beta_j^2}}{2}}}^{\mu}d\mu^\prime\left(\mu^{\prime2}-\frac{\beta_j^2}
{4\mu^{\prime2}}-
\alpha_j\right)^{3/2}\nonumber\\
&=&\frac{3(\alpha_j^2-\beta_j^2)}{16}\ln\frac{\sqrt{
\mu^2-\alpha_j/2+\sqrt{\alpha_j^2+\beta_j^2}/2}+
\sqrt{\mu^2-\alpha_j/2-\sqrt{\alpha_j^2+\beta_j^2}/2}}{\sqrt{
\mu^2-\alpha_j/2+\sqrt{\alpha_j^2+\beta_j^2}/2}-\sqrt{\mu^2-\alpha_j/2-
\sqrt{\alpha_j^2+\beta_j^2}/2}}\nonumber\\
&&+\frac{3\alpha_j|\beta_j|}{4}\arctan\sqrt{\frac{(\sqrt{\alpha_j^2+\beta_j^2}-\alpha_j)
(\mu^2-\sqrt{\alpha_j^2+\beta_j^2}/2-\alpha_j/2)}{(\sqrt{\alpha_j^2+\beta_j^2}+\alpha_j)
(\mu^2+\sqrt{\alpha_j^2+\beta_j^2}/2-\alpha_j/2)}}\nonumber\\
&&+\frac{\mu^2}{4}\sqrt{\mu^4-\alpha_j\mu^2-\beta_j^2/4}
-\frac{5\alpha_j}{8}\sqrt{\mu^4-\alpha_j\mu^2-\beta_j^2/4}
+\frac{\beta_j^2}{8}\frac{\sqrt{\mu^4-\alpha_j\mu^2-\beta_j^2/4}}{\mu^2}.\nonumber\\
\end{eqnarray}

Now let us turn to the calculation of ${\cal P}(\mu)|_{\mu=0}$. The rainbow-ladder approximation of DSE is the stationary point equation for the CJT effective action [28] which, evaluated at this stationary point, is [29]
\begin{eqnarray}
{\cal P}(\mu)|_{\mu=0} &=& 2N_cN_f\int\frac{d^4p}{(2\pi)^4}\left\{\ln\left[\frac{A^2(p^2)p^2+B^2(p^2)}{p^2}\right]-\frac{p^2A(p^2)[A(p^2)-1]+B^2(p^2)}{p^2A^2(p^2)+B^2(p^2)}\right\} \nonumber \\
&=& -2N_cN_f\int\frac{d^4p}{(2\pi)^4} \bigg\{ \ln \bigg[ p^2 \bigg(p^2 \sigma_v^2(p^2)+\sigma_s^2(p^2)\bigg)\bigg]+1+p^2 \sigma_v (p^2) \bigg\},
\end{eqnarray}
where $G(p)=\frac{1}{i \gamma \cdot p A(p^2)+B(p^2)} \equiv i \gamma \cdot p \sigma_v (p^2)+\sigma_s(p^2)$ is the unrenormalized dressed quark propagator at $\mu=0$. From the model quark propagator (9) one obtains
\begin{equation}
\sigma_v(p^2)=-\sum_{j=1}^{n_P}\bigg(\frac{r_j}{p^2+m_j^2}+\frac{r_j}{p^2+m_j^{*2}}\bigg),~~~~~\sigma_s(p^2)=\sum_{j=1}^{n_P}\bigg(\frac{r_j
m_j}{p^2+m_j^2}+\frac{r_j m_j^*}{p^2+m_j^{*2}}\bigg).
\end{equation}
From Eqs. (20,21) and the parameters of the model quark propagator (9) listed in Table I, one can calculate ${\cal P}(\mu)|_{\mu=0}$.

The determination of the EOS of QCD is a longstanding problem in strong interaction physics. Lattice QCD calculations and phenomenological models try to pin down a useable EOS since two decades.
It is interesting to compare the EOS obtained in this paper to the EOS of QCD proposed in previous studies. Here we shall take one prominent example, the cold, perturbative EOS of QCD proposed by Fraga, Pisarski and Schaffner-Bielich in Ref. [30]. The pressure density to second order in $\alpha_s$ in the $\overline{MS}$ scheme obtained in Ref. [30] is quoted as follows
\begin{equation}
{\cal P}_{FPS}(\mu)=\frac{N_f\mu^4}{4\pi^2}\Bigg\{1-2\Big(\frac{\alpha_s}{\pi}\Big)-\Bigg[G+N_f \ln \frac{\alpha_s}{\pi}+\Big(11-\frac{2}{3}N_f \Big)\ln\frac{{\bar \Lambda}}{\mu}\Bigg]\Big(\frac{\alpha_s}{\pi}\Big)^2\Bigg\},
\end{equation}
where $G=G_0-0.536N_f+N_f \ln N_f$, $G_0=10.374 \pm 0.13$ and ${\bar \Lambda}$ is the renormalization subtraction point. The scale dependence of the strong coupling constant $\alpha_s({\bar \Lambda})$ is taken as
\[
\alpha_s({\bar \Lambda})=\frac{4\pi}{\beta_0 u}\Bigg[ 1-\frac{2\beta_1}{\beta_0^2}\frac{\ln(u)}{u}+\frac{4\beta_1^2}{\beta_0^4 u^2}\Bigg(\Big(\ln(u)-\frac{1}{2}\Big)^2+\frac{\beta_2\beta_0}{8\beta_1^2}-\frac{5}{4}\Bigg) \Bigg],
\]
where $u=\ln({\bar \Lambda}^2/\Lambda_{\overline{MS}}^2)$, $\beta_0=11-2N_f/3$, $\beta_1=51-19N_f/3$, and $\beta_2=2857-5033N_f/9+325N_f^2/27$. For $N_f=3$, $\Lambda_{\overline{MS}}=365~\mathrm{MeV}$. The only freedom in the model of Ref. [30] is the choice of the ratio ${\bar \Lambda}/\mu$, which is taken to be 2 in that reference. The perturbative EOS (22) is applicable only in the chirally symmetric phase, when the chemical potential $\mu$ is larger than $\mu_\chi$, the chiral phase transition point. A plot of our EOS and the EOS of Fraga, Pisarski and Schaffner-Bielich is given in Fig. 4. Here we note that when applying our EOS to the study of neutron star, owing to the boundary condition imposed on the surface of neutron star, the constant term ${\cal P}(\mu)|_{\mu=0}$ does not contribute to the mass-radius relation, so when comparing our EOS with that of Fraga, Pisarski and Schaffner-Bielich in Fig. 4, we do not consider this term. From Fig. 4 it can be seen that 
in the region of $\mu$ studied, the pressure density in the EOS of Fraga, Pisarski and Schaffner-Bielich and our EOSs (for the three parametrization of the model quark propagator) increases monotonically as $\mu$ increases. 
In large $\mu$ region the EOS of Fraga, Pisarski and Schaffner-Bielich and our EOSs show qualitatively similar behaviors: all of them tend to the free quark gas result as $\mu$ tends to infinity. Compared with the EOS of Fraga, Pisarski and Schaffner-Bielich, our EOSs tend somewhat more rapidly to the free quark gas result. When $\mu$ is less than about $1.8~\mathrm{GeV}$,  the EOS of Fraga, Pisarski and Schaffner-Bielich and our EOSs begin to show appreciable difference: the pressure density given in our EOSs is much lower than that given in the EOS of Fraga, Pisarski and Schaffner-Bielich. Here we note that the form of our EOS depends strongly on the model quark propagator at zero $\mu$ one chooses. The parameters of the model quark propagator (9)
are determined by numerical fitting [25] and there exist some arbitrariness in this process. In further researches we will apply our EOS to the study of neutron star and we hope that by comparing theoretical results with observational results of neutron star we can have further constraints to the parameters of our model quark propagator.

\begin{figure}[ht]
\centering\includegraphics[width=13cm]{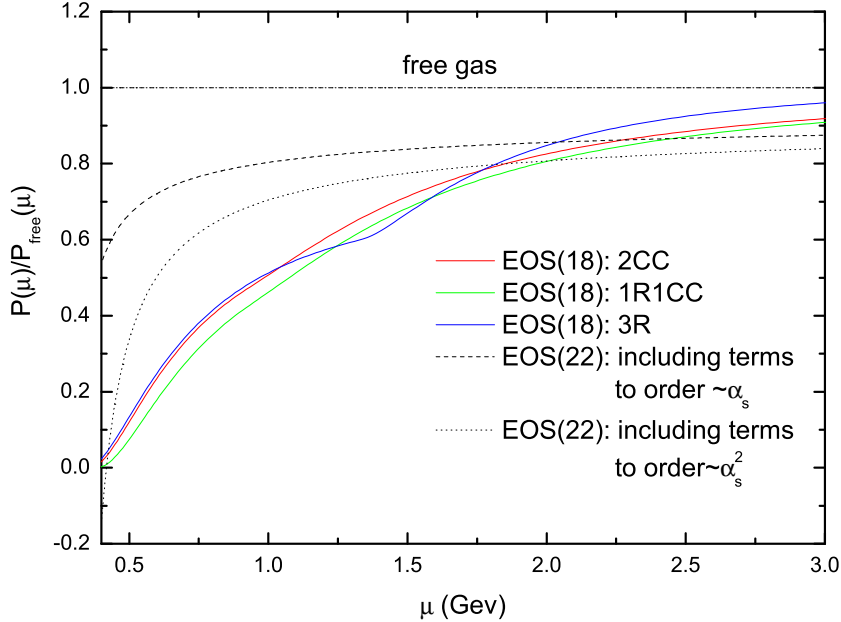}\\[-1cm]
\caption{The pressure, relative to the free quark gas pressure
${\cal P}_{\mathrm free}=N_cN_f\mu^4/(12\pi^2)$, in our EOS (18) and
the EOS (22) of Fraga, Pisarski and Schaffner-Bielich}\label{fig4}
\end{figure}

To summarize, in the study of strongly interacting matter at finite temperature and/or finite density, the knowledge of the partition function of QCD determines all the thermodynamic properties of the system. In this paper, we try to give a direct method for calculating the partition function, and hence the EOS of QCD at finite chemical potential and zero temperature. In this method the quark-number density $\rho(\mu)$ is expressed in terms of the dressed quark propagator at finite chemical potential $G[\mu](p)$, and the pressure density ${\cal P}(\mu)$ is given by the integration of $\rho(\mu)$ plus an additive constant, whose physical meaning is the pressure density at zero chemical potential ${\cal P}(\mu)|_{\mu=0}$. Because of the difficulty of calculating $G[\mu](p)$ and ${\cal P}(\mu)|_{\mu=0}$ from first principles of QCD, one has to resort to nonperturbative QCD models when applying our method. In this paper we adopt one nonperturbative QCD model, the rainbow-ladder approximation of the Dyson-Schwinger approach. In order to obtain an EOS with explicit analytical form, we choose the meromorphic model quark propagator proposed in Ref. [25]. Using the general result proved in the framework of the rainbow-ladder approximation of the DS approach in Refs. [17,18], $G[\mu](p)$ is obtained from this model quark propagator. From this the quark-number density $\rho(\mu)$ is calculated, which is found to differ significantly from the Fermi distribution of free quark theory. Physically this is a consequence of dynamical chiral symmetry breaking and confinement in the low energy region. It is found that when $\mu$ is below a critical value $\mu_0$ ($\mu_0=351~\mathrm{MeV}$, $377~\mathrm{MeV}$ and $341~\mathrm{MeV}$ for the 2CC, 1R1CC and 3R parametrization, respectively), the quark-number density vanishes identically. This feature agrees with the general conclusion in Ref. [27]. The value $\mu_0$ obtained here is almost of the same order of magnitude as the estimate made in [27] ($307~\mathrm{MeV}$). The constant ${\cal P}(\mu)|_{\mu=0}$ is also self-consistently calculated using the rainbow-ladder approximation of the DS approach. From these the full analytic expression of the EOS of QCD at finite $\mu$ and zero $T$ is obtained (apart from the constant term ${\cal P}(\mu)|_{\mu=0}$ which can in principle be calculated from the CJT effective action). A comparison between our EOS and one prominent example of the EOS of QCD, the cold, perturbative EOS of QCD proposed by Fraga, Pisarski and Schaffner-Bielich in [30] is made.

\vspace*{0.8cm}

\noindent{\large \bf Acknowledgments}

This work was supported in part by the National Natural Science Foundation of China (under Grant No 10575050) and the Research Fund for the Doctoral Program of Higher Education (under Grant No 20060284020).

\vspace*{0.8cm}
\noindent{\large \bf References}
\begin{description}
\item{[1]} J.I. Kapusta, Finite-temperature field theory (Cambridge University Press, Cambridge, 1989).
\item{[2]} M.le Bellac, Thermal Field Theory (Cambridge University Press, Cambridge, 1996).
\item{[3]} Z. Fodor and S. Katz, Phys. Lett. {\bf B 534}, 87 (2002).
\item{[4]} M. D'Elia and M.P. Lombardo, Phys. Rev. {\bf D 67}, 014505 (2003).
\item{[5]} C.R. Allton, S. Ejiri, S. J. Hands, O. Kaczmarek, F. Karsch, E. Laermann and C. Schmidt, Phys. Rev. {D 68}, 014507 (2003).
\item{[6]} R.V. Gavai and S. Gupta, Phys. Rev. {\bf D 68}, 034506 (2003); R.V. Gavai and S. Gupta, ibid. {\bf D 72}, 054006 (2005).
\item{[7]} S. Gupta and R. Ray, Phys. Rev. {\bf D 70}, 114015 (2004).
\item{[8]} M. He, W.M. Sun, H.T. Feng, and H.S. Zong, J. Phys. G: Nucl. Part. Phys. {\bf 34}, 2655 (2007); W.M. Sun and H.S. Zong, Int. J. Mod. Phys. {\bf A 22}, 3201 (2007).
\item{[9]} C.R. Allton et al., Phys. Rev. {\bf D 71}, 054508 (2005).
\item{[10]} F. $\ddot{O}$zel, Nature {\bf 445}, 1115 (2006).
\item{[11]} M. Alford, et al., Nature {\bf 445}, E7 (2007).
\item{[12]} D. Nickel, J. Wambach and R. Alkofer, Phys. Rev. {\bf D 73}, 114028 (2006).
\item{[13]} C.D. Roberts and A.G. Williams, Prog. Part. Nucl. Phys. {\bf 33}, 477 (1994), and references therein.
\item{[14]} C.D. Roberts and S.M. Schmidt, Prog. Part. Nucl. Phys. {\bf 45S1}, 1 (2000), and references therein.
\item{[15]} P. Maris and C.D. Roberts, Int. J. Mod Phys. E {\bf 12}, 297 (2003).
\item{[16]} R. Alkofer and L.von Smekal, Phys. Rept. {\bf 353}, 281 (2001); C.S. Fischer and R. Alkofer, Phys. Rev. D {\bf 67}, 094020 (2003), and references therein.
\item{[17]} H.S. Zong, L. Chang, F.Y. Hou, W.M. Sun and Y.X. Liu, Phys. Rev. {\bf C 71}, 015205 (2005); F.Y. Hou, L. Chang, W.M. Sun, H.S. Zong and Y.X. Liu, Phys. Rev. {\bf C 72}, 034901 (2005).
\item{[18]} H.T. Feng, F.Y. Hou, X. He, W.M. Sun and H.S. Zong, Phys. Rev. {\bf D 73}, 016004 (2006); H.T. Feng, W.M. Sun, D.K He, and H.S. Zong, Phys. Lett. {\bf B 661}, 57 (2008).
\item{[19]} Y. Taniguchi and Y. Yoshida, Phys. Rev. D {\bf 55}, 2283 (1997).
\item{[20]} D. Blaschke, C.D. Roberts and S. Schmidt, Phys. Lett. {\bf B 425}, 232 (1998).
\item{[21]} P. Maris, C.D. Roberts and P.C. Tandy, Phys. Lett. {\bf B 420}, 267 (1998).
\item{[22]} A. Bender, W. Detmold and A.W. Thomas, Phys. Lett. {\bf B 516}, 54 (2001).
\item{[23]} O. Miyamura, S. Choe, Y. Liu, T. Takaishi, and A. Nakamura, Phys. Rev. {\bf D 66}, 077502 (2002).
\item{[24]} See, e.g., M.J. Ablowitz and A.S. Fokas, {\it Complex Variables, Introduction and Applications, 2nd ed.} (Cambridge University Press, 2003) p. 122. 
\item{[25]} R. Alkofer, W. Detmold, C.S. Fischer, and P. Maris, Phys. Rev. D {\bf 70}, 014014 (2004).
\item{[26]} M.S. Bhagwat, M.A. Pichowsky, and P.C. Tandy, Phys. Rev. {\bf D 67}, 054019 (2003).
\item{[27]} M. A. Halasz, A. D. Jackson, R. E. Shrock, M. A. Stephanov and J. J. M. Verbaarschot, Phys. Rev. {\bf D 58}, 096007 (1998).
\item{[28]} J. M. Cornwall, R. Jackiw and E. Tomboulis, Phys. Rev. {\bf D 10}, 2428 (1974).
\item{[29]} K. Stam, Phys. Lett. {\bf B 152}, 238 (1985).
\item{[30]} E. S. Fraga, R. D. Pisarski, and J. Schaffner-Bielich, Phys. Rev. {\bf D 63}, 121702(R) (2001); Nucl. Phys. {\bf A 702}, 217c (2002).

\end{description}

\end{document}